\documentclass[a4paper]{article}

\pdfoutput=1

\usepackage[english]{babel}
\usepackage[utf8x]{inputenc}
\usepackage[T1]{fontenc}

\usepackage[a4paper,top=3cm,bottom=2cm,left=3cm,right=3cm,marginparwidth=1.75cm]{geometry}

\usepackage{amsmath}
\usepackage{mathtools}
\usepackage{graphicx}
\usepackage{listings}
\usepackage[colorinlistoftodos]{todonotes}
\usepackage{url}
\usepackage[colorlinks=true, allcolors=blue,breaklinks]{hyperref}
\usepackage{cleveref}

\title{On the computation of Shannon Entropy from Counting Bloom Filters}
\author{Michael Cochez}

\begin{document}
\maketitle

\section{Discrete Entropy Estimate}

In this write-up a method for computing the naive plugin estimator of discrete entropy from a counting Bloom filter will be presented.
The entropy estimate, or in the rest of this text just the entropy, we will compute is:

\begin{equation}
\hat{H} = - \sum_{k=1}^K \hat{p}_k \log \hat{p}_k,\label{Hplugin1}
\end{equation}

Where $\hat{p}_k$ is the fraction of elements in the sample having value $k$ and $0 * \log(0) = 0$ by convention.
This will underestimate the entropy in most cases, for mitigation, see \url{http://www.nowozin.net/sebastian/blog/estimating-discrete-entropy-part-1.html}.

\section{Counting Bloom filters}

A Bloom filter~\cite{bloom1970space} is used to perform set membership testing without having the actual data in a set available.
The test can give false positives (outputs that the tested element is in the set while it is not), but never false negatives (if the test tells the element is not in the set, it definitely is not).
The data structure consists of an array, which is initially set to all zeros, and a set of hash functions. 
The outcome range of the hash functions is exactly as large as the length of the array.
To add an element to the structure (i.e., register it as en element of the set), the element is hashed with each of the hash functions and the bits at the indexes which corresponds to the outcome of the hash function applications is set to 1.

To test whether an element is in the set, the same hash functions are applied on the element and each of the bits at indexes corresponding to their outcomes is checked.
If any of these bits is set to 0, then is concluded that the element was not in the set.
This is correct because if it had been there, all bits would have been set to 1 at an earlier point.
If all these bits are set to 1, it is concluded that the element \emph{might} be in the set.
Either, the element is really in the set and hence its bits were set to one at an earlier point, or all of these bits were set to 1 by the application of the hash functions on other elements which are in the set. In the latter case we consider the outcome a false positive.

A drawback of Bloom filters is that it is impossible to remove any elements from it.
One might naively think that one could remove an element by setting all bits corresponding to the outcome of the hash functions to 0.
However, this might violate the properties described above in case any of these bits was also set to 1 by any hash function application on any other element which is still in the set.
To overcome this issue Li Fan et al.~\cite{fan2000summary} introduced counting Bloom filters.

A counting Bloom filter works like a normal Bloom filter, but instead of maintaining a bit array, an array with integers, initially filled with zeros, is used.
When an element is added to the filter, the hash functions are computed and instead of setting the corresponding bits to 1, the count at these indexes is incremented with 1.
The test for membership will conclude that the element is not in the set in case the counts at any of the hash function outcome indexes is 0.

\section{Estimating Discrete Entropy from a Counting Bloom Filter}

The naive plugin estimator of discrete entropy described above can be computed for a collection of observations.
Now, if this set is not available, but a counting Bloom filter for that set is, we can still compute an approximation of this entropy.

Given the counting Bloom filter array $BF$, and assuming $m$ hash functions were used to insert $c$ elements. 
Then, we can compute the following formula on the numbers in the Bloom filter array:

\begin{align}
\shortintertext{Uncorrected Bloom filter entropy}
\hat{H}_{uncorreted BF} = - \sum_{k=1}^{|BF|} \hat{p}_k \log \hat{p}_k,\label{uncorrectedEntropyBF} 
\end{align}

This is pretty much the same as \cref{Hplugin1}, but this time we will set $\hat{p}_k$ to $BF[k] / c$, and not $BF[k] / (m*c)$, which would account for the total count of the numbers in the Bloom filter.
Assuming that there are no collision in the Bloom filter, this computation will be equal to the entropy of \cref{Hplugin1}, except for the fact that each element which was added to the Bloom filter has $m$ occurrences in the Bloom filter array.
What this means is that in the summation in \cref{uncorrectedEntropyBF} each unique entry is contributing $m$ times what it was supposed to.
Hence, to compute the correct value, we need to scale this sum by a factor $1/m$ and we obtain:
\begin{align}
\shortintertext{Bloom filter entropy}
\hat{H}_{BF} = - \frac{1}{m} \sum_{k=1}^{|BF|} \hat{p}_k \log \hat{p}_k,\label{coequationrrectedEntropyBF} 
\end{align}
with $\hat{p}_k$ as $BF[k] / c$.

This computation will always result in a value lower than that of the estimator of discrete entropy. 
The reason is that each collision in the bloom filter will reduce the entropy.
Hence, if it is possible to check multiple bloom filters, the highest obtained result must be used for the final outcome.

\subsection{Further Possible Corrections}

To correct this value, one needs to account for collisions in the Bloom filter.
In general, one cannot know that a collision happened, but there are some cases where one can be certain.
For example, if there are 3 hash functions and the Bloom filter contains (only) the counts $[1,1,3,2,2]$ (leaving possible zeros out), then one can be sure that the $3$ is caused by a collision.
But, we can still not be completely certain whether there was one element of say type A (causing the ones and the collision) and two elements of type B (causing the twos and the collision).
It could also be the case that there is one element of types A, B, and C, and that more collisions happened to lead to this result.
The second scenario is, however, much less likely, especially assuming that collisions in Bloom filters are rather rare events.
In the second scenario we would indeed need more collisions than in the first one.
So, in general, one could attempt to correct the approximation in the previous section by computing the entropy for the most likely scenario.
However, for the moment, we have not found a general solution for finding this most likely scenario (or perhaps the expected value over all possible scenarios). 
From some initial though experiments is seems that finding this general solution will be very complex computationally.
Further analysis is left for future research or version of this work.

\bibliographystyle{alpha}
\bibliography{refs}

\begin{thebibliography}{FCAB00}

\bibitem[Blo70]{bloom1970space}
Burton~H Bloom.
\newblock Space/time trade-offs in hash coding with allowable errors.
\newblock {\em Communications of the ACM}, 13(7):422--426, 1970.

\bibitem[FCAB00]{fan2000summary}
Li~Fan, Pei Cao, Jussara Almeida, and Andrei~Z Broder.
\newblock Summary cache: a scalable wide-area web cache sharing protocol.
\newblock {\em IEEE/ACM transactions on networking}, 8(3):281--293, 2000.

\end{thebibliography}

\appendix

\section{Java code}

	\begin{description}
		\item[$BF$] The contents of the counting Bloom filter. Zeros can be left out. 
		\item[$m$] The number of hash functions used to fill the filter.
	\end{description}
\begin{lstlisting}[language=Java]
public static double entropyEstimate(Iterable<Short> BF, int m) {
	long totalCountBF = 0;
	for (short s : BF) {
		totalCountBF += s;
	}
	double normalizedCount = totalCountBF / m;
	
	double entropy = 0;
	for (short s : BF) {
		if (s != 0) {
			double P = s / normalizedCount;
			entropy += P * DoubleMath.log2(P);
		}
	}
	return -(1 / (double) m) * entropy;
}
\end{lstlisting}

\end{document}